\documentclass[%
prb,twocolumn,
superscriptaddress,
amsmath,amssymb,
 aps,
]{revtex4-2}

\usepackage{graphicx}
\usepackage{dcolumn}
\usepackage{bm}
\usepackage[mathlines]{lineno}

\usepackage{physics}
\usepackage{amsmath}
\usepackage{blkarray}
\usepackage{xcolor}
\usepackage{natbib}
\usepackage[pagebackref=false,colorlinks,linkcolor=blue,citecolor=blue,urlcolor=blue]{hyperref}
\usepackage{soul}
\graphicspath{{Figures/}}
\newcommand{\be}{\begin{equation}}

\newcommand{\ee}{\end{equation}}
\newcommand{\bea}{\begin{eqnarray}}
\newcommand{\eea}{\end{eqnarray}}

\renewcommand{\vec}[1]{\mathbf{ #1}}

\renewcommand{\tilde}{\widetilde}

\begin{document}

\preprint{APS/123-QED}

\title{Finite-\textit{q} photon-drag shift current in two-dimensional massive chiral Dirac fermions}

\author{Rofii}
\email{rofiibahrudin@gmail.com}
\affiliation{
Research Center for Physics, National Research and Innovation Agency, Tangerang Selatan 15314, Indonesia.
}%
\affiliation{
Department of Physics, Universitas Airlangga, Surabaya 60115, Indonesia
}%

\author{Eddwi Hesky Hasdeo}
\email{eddw001@brin.go.id}
\affiliation{
Research Center for Physics, National Research and Innovation Agency, Tangerang Selatan 15314, Indonesia.
}

\date{\today}

\begin{abstract}
We investigate the photon-drag shift current in an isotropic single-valley two-dimensional massive chiral Dirac model with chirality index $J=1,2,3$ by directly evaluating the full finite-$q$ non-vertical response beyond the perturbative small-$q$ regime.
Our central result is that chirality qualitatively reorganizes the sign topology of the finite-$q$ photocurrent $\vec j(\vec q)$.
For $J=1$, the photocurrent remains broadly positive, whereas higher-chirality sectors ($J \ge 2$) generically develop internal zero-current contours and sign reversals within the kinematically allowed region.
We further show that the photocurrent is symmetry-constrained to be purely transverse, $\mathbf{j}(\mathbf{q}) \propto \hat{\mathbf{z}}\times\mathbf{q}$, and vanishes in the strictly vertical-transition limit $q=0$ in centrosymmetric systems.
Pauli blocking further shapes the response by selecting the active portion of
the resonance contour, while its extinction at large \(\Delta\) or \(q\) follows
from a simple kinematic cutoff.
These results establish the isotropic massive chiral Dirac problem as a symmetry-controlled benchmark for chirality-dependent finite-$q$ shift currents.
\end{abstract}

\maketitle


\section{\label{sec:intro} Introduction}

\begin{figure}
\includegraphics[width=\columnwidth]{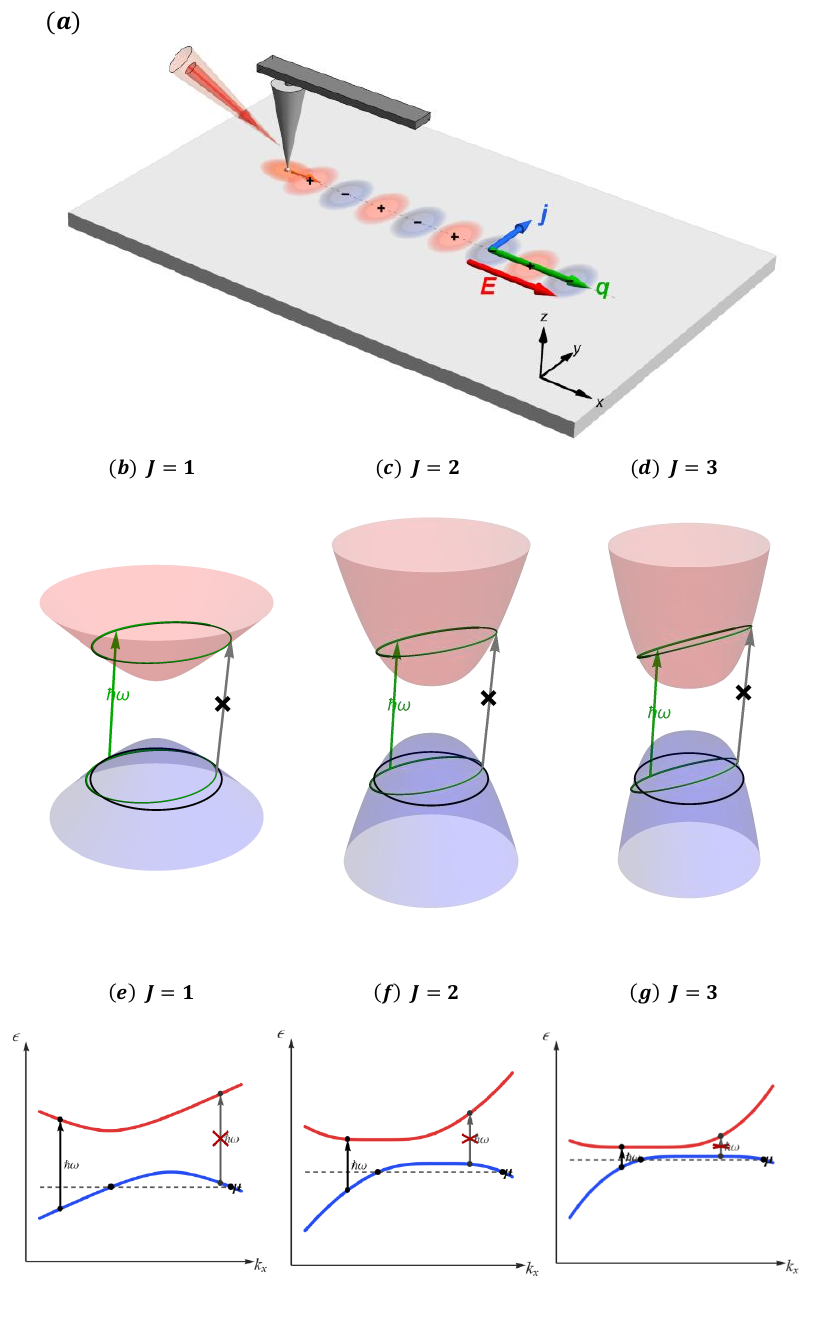}
\caption{\label{fig:transition}
(a) Finite-$q$ driving geometry considered in this work. A sharp metallic tip locally converts the incident electromagnetic field into a propagating in-plane plasmonic wave with wave vector $\mathbf q$. The electric field is taken parallel to the propagation direction, $\mathbf E \parallel \mathbf q$; for $\mathbf q = q\hat{\mathbf x}$, the dc photocurrent is purely transverse, $\mathbf j(\mathbf q)\propto \hat{\mathbf z}\times\mathbf q$.
(b--d) Schematic illustration of finite-$q$ non-vertical interband transitions in the massive chiral Dirac dispersion for (b) $J=1$, (c) $J=2$, and (d) $J=3$, shown in the original-band picture. The red and blue surfaces denote the conduction and valence bands, respectively. The green contours mark the energy-conserving contour (EC), whereas the black contours indicate the chemical-potential boundary. The green arrow denotes an allowed transition, while the gray arrow denotes a Pauli-blocked one. 
(e--g) Equivalent representation of the same finite-$q$ process for (e) $J=1$, (f) $J=2$, and (g) $J=3$, viewed as a vertical transition between momentum-shifted valence and conduction bands. In this representation, the finite-$q$ transition $|u_v(\mathbf{k}-\mathbf{q}/2)\rangle \to |u_c(\mathbf{k}+\mathbf{q}/2)\rangle$ is visualized as a vertical excitation between the shifted bands; the dashed line indicates the chemical potential $\mu$. 
}
\end{figure}

The bulk photovoltaic effect (BPVE) is the generation of a dc photocurrent in a spatially homogeneous crystal under uniform illumination, without requiring p--n junctions or built-in electric fields, and constitutes a central nonlinear optical response of solids \cite{Fridkin2001,Dai2023,Glass1974,Sturman1992}.
Among its microscopic mechanisms, the shift current is a coherent interband contribution that can be understood as a real-space displacement of the photoexcited electronic wavepacket governed by the quantum geometry of Bloch states \cite{Baltz1981,Sipe2000,Morimoto2016}.
In the usual vertical-transition limit, however, inversion symmetry applies stringent constraints on the second-order conductivity tensor, so that the shift-current response vanishes in centrosymmetric crystals \cite{Ahn2020,Young2012}.

Recent theoretical work has shown that this conclusion can be circumvented when optical excitation involves non-vertical interband transitions from the valence, $v$, to the conduction, $c$, bands with a finite photon momentum transfer $\mathbf{q}$, schematically $\lvert v,\mathbf{k}-\mathbf{q}/2\rangle \rightarrow \lvert c,\mathbf{k}+\mathbf{q}/2\rangle$ \cite{Likun2021,Nagaosa2025}.
Although the bare photon momentum is ordinarily small, finite-$q$ effects can be strongly enhanced in optical settings with large effective wavevectors, such as polaritonic confinement \cite{Xiong2022,Durach2009} or structured beams \cite{Gunyaga2023}.
A particularly relevant realization for the present work is near-field excitation by a sharp metallic tip, as in scattering-type scanning near-field optical microscopy (s-SNOM), where the incident electromagnetic field can be locally converted into a propagating in-plane plasmonic wave carrying a large in-plane momentum, as illustrated schematically in Fig.~\ref{fig:transition}(a).
Despite this progress, the present theoretical understanding remains limited in two important respects.
First, most analyses are effectively anchored to the perturbative small-$q$ regime.
Second, explicit demonstrations have so far been concentrated in a relatively narrow class of two-dimensional effective models, most notably BHZ-type Hamiltonians \cite{Likun2021,Nagaosa2025}.
It therefore remains unresolved how the full finite-$q$ photon-drag shift current is organized in a minimal model that remains both centrosymmetric and rotationally symmetric, and how its photocurrent response varies with the chirality index $J$ beyond the perturbative small-$q$ regime.

A natural setting in which to address these questions is provided by the family of massive chiral Dirac Hamiltonians, which preserve a minimal two-band structure while allowing the chirality index $J$ to be tuned.
Here, $J$ controls the winding of the internal pseudospin texture, rather than merely the radial shape of the band dispersion.
These models therefore furnish a controlled benchmark for isolating the role of pseudospin winding in the finite-$q$ shift-current integrand before additional lattice anisotropy, trigonal warping, or valley-resolving ingredients are introduced.
They are relevant not only to effective descriptions of chirally stacked or rhombohedral multilayer graphene \cite{Hongkimin2008,Koshino2009,Zhang2011,McCann2013}, where higher-order chiral Dirac fermions arise naturally, but also to the low-energy chiral sectors underlying higher-Chern insulating phases in related systems \cite{Zhang2011,Liu2025,Winterer2024,Han2024, Sha2024}.

In this work, we use the isotropic single-valley massive chiral Dirac model
with \(J=1,2,3\) as a minimal setting to isolate the role of pseudospin winding
in finite-\(q\) photon-drag shift currents [see Fig.~\ref{fig:transition}(b--d)].
By evaluating the full finite-$q$ non-vertical shift-current expression, rather than truncating the response at leading order in $q$, we systematically map its dependence on the chemical potential $\mu$, band gap $\Delta$, optical frequency $\omega$, and photon momentum $\mathbf{q}$.
We show that the photocurrent is constrained to be transverse to the photon
momentum and vanishes in the strictly vertical-transition limit.
The main finding is that chirality changes the internal sign structure of the finite-\(q\) response: the \(J=1\) sector remains broadly positive, whereas the
\(J\ge2\) sectors develop internal zero-current contours and sign reversals
within the kinematically allowed region.
These sign changes are controlled by
the interplay between resonance selection, Pauli blocking, and the geometric shift vector, while the final extinction of the response is set by a simple kinematic cutoff.
These results establish the isotropic massive chiral Dirac problem as a symmetry-controlled benchmark for chirality-dependent finite-$q$ shift currents.

\section{\label{sec:method} Model and Formalism}

We consider an isotropic single-valley two-band model of a gapped chiral Dirac fermion \cite{Hongkimin2008}, described by the effective Hamiltonian
\begin{equation}
H(\mathbf{k})=\varepsilon_0 \left[ \left( \frac{k_-}{k_0} \right)^J \sigma_+ + \left( \frac{k_+}{k_0} \right)^J \sigma_- \right] + \Delta \sigma_z - \mu\sigma_0 \, ,
\label{eq:Hk_chiral}
\end{equation}
where \(k_\pm=k_x \pm i k_y\), \(\sigma_\pm=\frac{1}{2}\left( \sigma_x \pm i \sigma_y \right)\), \(\Delta\) is the band gap (mass term), and \(\mu\) is the chemical potential.
The material-dependent parameters \(k_0\) and \(\varepsilon_0\) set the characteristic momentum and energy scales, respectively.
The integer \(J=1,2,3\) is the chirality index, which controls the winding of the pseudospin texture in momentum space.
The corresponding eigenenergies are
\begin{equation}
\varepsilon_\pm(\mathbf{k})
=
\varepsilon_0
\left[
\pm \sqrt{\tilde{k}^{\,2J}+\tilde{\Delta}^2}
-
\tilde{\mu}
\right],
\label{eq:energy}
\end{equation}
where $\tilde{k}=|\mathbf{k}|/k_0$, $\tilde{q}=q/k_0$, $\tilde{\Delta}=\Delta/\varepsilon_0$,
$\tilde{\mu}=\mu/\varepsilon_0$, and $\tilde{\omega}=\hbar\omega/\varepsilon_0$.
In this minimal model, \(\varepsilon_0\) defines the characteristic energy scale of the chiral dispersion, while \(\Delta\) acts as a mass term that opens the gap.
In more realistic effective descriptions, such as rhombohedral multilayer graphene, \(\varepsilon_0\) is typically set by intralayer--interlayer hopping ratio, whereas \(\Delta\) may be associated with an interlayer potential difference induced by a perpendicular electric field.

Throughout this work, we restrict attention to the isotropic limit and neglect trigonal warping and other higher-order anisotropic terms, so that the model remains rotationally symmetric in the \(k_x\)--\(k_y\) plane.
In this high-symmetry setting, the ordinary \(q=0\) shift current is symmetry-forbidden, but we will show that a finite-\(q\) photon-drag shift current can nevertheless emerge from non-vertical interband transitions.

To describe this effect, we directly evaluate the finite-$q$ generalization of the non-vertical shift-current formula.
For a photon (or polariton) carrying in-plane momentum $\mathbf q$, an interband transition connects an initial valence-band state $|u_v(\mathbf k-\mathbf q/2)\rangle$ to a final conduction-band state $|u_c(\mathbf k+\mathbf q/2)\rangle$. The same finite-$q$ process may be viewed from two equivalent perspectives.
In the original-band picture [Fig.~\ref{fig:transition}(b--d)], it appears as a non-vertical interband transition connecting states separated by the in-plane momentum transfer $\mathbf q$.
Equivalently, in the momentum-shifted representation [Fig.~\ref{fig:transition}(e--g)], the same process is visualized as a vertical transition between shifted valence and conduction bands.
In either representation, a finite-$q$ transition contributes only when it satisfies both energy conservation and Pauli blocking: the initial valence state must be occupied, while the final conduction state must remain empty.

For linearly polarized light, the photon-drag shift current is given by \cite{Likun2021}
\begin{equation}
\mathbf{j}(\mathbf{q})
= C \int \frac{d^2\vec k}{(2\pi)^2} \rho(\mathbf{k},\mathbf{q})\,\left|\nu(\mathbf{k},\mathbf{q})\right|^{2}\,\mathbf{r}(\mathbf{k},\mathbf{q}),
\label{eq:pdsc}
\end{equation}
where $\rho(\mathbf{k},\mathbf{q}) = \delta\!\left(\omega_{+,\mathbf{k}+\mathbf{q}/2}-\omega_{-,\mathbf{k}-\mathbf{q}/2}-\omega\right)\Bigl[f\!\left(\varepsilon_{-,\mathbf{k}-\mathbf{q}/2}\right)$
$-f\!\left(\varepsilon_{+,\mathbf{k}+\mathbf{q}/2}\right)\Bigr]$
is the phase-space factor encoding both Pauli blocking and energy conservation.
It therefore selects the optically allowed energy-conserving contour (EC) in momentum space.
The prefactor \(C\) collects the dependence on the electric-field amplitude and optical frequency; for linearly polarized light in our convention,
\(C=e(\pi/2)(eE/\hbar\omega)^2\).
For light polarized along the \(x\) axis, the interband velocity matrix element is defined as
\begin{equation}
\nu(\mathbf{k},\mathbf{q})
=
\biggl\langle u_c\!\left(\mathbf{k}+\frac{\mathbf{q}}{2}\right)\bigg|
\frac{1}{\hbar}\frac{\partial H}{\partial k_x}
\bigg|u_v\!\left(\mathbf{k}-\frac{\mathbf{q}}{2}\right)\biggr\rangle.
\label{eq:v-matrix}
\end{equation}
The real-space displacement associated with a non-vertical transition is encoded in the generalized shift vector,
\begin{equation}
\mathbf{r}(\mathbf{k},\mathbf{q})
=
\mathbf{A}_c(\mathbf{k}+\mathbf{q}/2)
-\mathbf{A}_v(\mathbf{k}-\mathbf{q}/2)
-\nabla_{\mathbf{k}}\arg\!\bigl[\nu(\mathbf{k},\mathbf{q})\bigr],
\label{eq:sv}
\end{equation}
where $\mathbf{A}_n(\mathbf k)=
i\langle u_n(\mathbf k)|\nabla_{\mathbf k}u_n(\mathbf k)\rangle,$ $n=c,v$, are the Berry connections of the corresponding bands.

For the present geometry, illustrated in Fig.~\ref{fig:transition}(a), the in-plane momentum is taken as $\mathbf q=q\,\hat{\mathbf x}$, as may arise from a propagating plasmonic wave locally launched by a sharp metallic tip.
The electric field is taken parallel to the propagation direction, $\mathbf E\parallel\mathbf q$, so that for $\mathbf q=q\,\hat{\mathbf x}$ the light is polarized along $\hat{\mathbf x}$.
As shown below, this geometry constrains the dc photocurrent to be purely transverse, as summarized by Eq.~\eqref{eq:symmetryform}.
Accordingly, for $\mathbf q=q\,\hat{\mathbf x}$, the current flows along $\hat{\mathbf y}$.

The present analysis is intentionally restricted to the isotropic single-valley limit, in which the role of chirality can be isolated most transparently.
For a time-reversal-related two-valley completion of the isotropic model, however, the second valley does not cancel the response obtained from Eq.~\eqref{eq:pdsc}.
Instead, the two valley contributions are identical. 
Indeed, if \(K'\) is related to \(K\) by time reversal, the corresponding currents satisfy (see Appendix~\ref{app:valley-relation})
\begin{equation}
\mathbf j_{K'}(\mathbf q)=-\mathbf j_K(-\mathbf q),
\label{eq:valeysrelation}
\end{equation}
where the additional minus sign follows from the time-reversal odd character of the current operator.

\section{Results and Discussion}
\label{sec:results}

\subsection{Symmetry and kinematic framework of the finite-\textit{q} response}
\label{sec:results-A}

Before turning to the detailed parameter dependence, we first establish the two general constraints that organize the finite-$q$ photon-drag shift current in the present isotropic single-valley geometry.
As summarized in Fig.~\ref{fig:transition}, the finite-\(q\) process is a non-vertical interband transition that can equivalently be viewed in the shifted-band representation.
The resulting response is organized by two constraints: symmetry fixes the current direction, while resonance kinematics and Pauli blocking determine whether the transition contributes.

The symmetry constraint can be established analytically.
For $\mathbf q \parallel \hat{x}$ and $x$-polarized light, the longitudinal part of the finite-$q$ shift-current integrand,
$\rho(\mathbf{k},\mathbf{q})\,|\nu(\mathbf{k},\mathbf{q})|^{2}\,\mathbf{r}(\mathbf{k},\mathbf{q})$,
is odd under the mirror operation $(k_x,k_y)\to(k_x,-k_y)$, whereas the transverse part is even.
After integration over momentum, the longitudinal photocurrent therefore vanishes identically, while the transverse component remains finite.
The dc current for the present model must thus take the  form (see Appendix~\ref{app:symmetryform})
\begin{equation}
\mathbf{j}(\mathbf{q})
=
\alpha(\omega,\Delta,\mu,|\mathbf{q}|)
\,\hat{\mathbf{z}}\times \mathbf{q},
\label{eq:symmetryform}
\end{equation}
where $\alpha$ is a scalar response function.
It follows immediately that the response is odd under photon-momentum reversal, $\mathbf{j}(-\mathbf{q})=-\mathbf{j}(\mathbf{q})$, and vanishes in the strictly vertical-transition limit, $\mathbf{j}(\mathbf{q}=0)=0$.
A finite in-plane photon momentum is therefore not merely a quantitative correction, but a necessary prerequisite for activating the transverse shift current in the present setting.

The second general constraint is kinematic.
A nonzero response requires the existence of an energy-conserving contour for the non-vertical interband transition.
Only after such a contour exists can Pauli blocking select which portion of it remains active.
The resonance condition leads directly to the kinematic cutoff (see Appendix~\ref{app:kinematic_cutoff})
\begin{equation}
\tilde\omega^2 \geq \left(2\tilde{\Delta}\right)^2 + \left(\frac{\tilde{q}^J}{2^{J-1}}\right)^2,
\label{eq:kin}
\end{equation}
which defines the outer boundary beyond which the resonance contour disappears and the photocurrent must vanish.
As shown below, Eq.~\eqref{eq:kin} reappears as the outer extinction boundary of the active region in the global phase diagram of Fig.~\ref{fig:qomega}.

These two constraints fix the broad structure of the finite-\(q\) response: symmetry enforces a transverse current, while resonance kinematics determines
whether a response can exist.
The remaining question is how Pauli blocking and the chirality-dependent integrand organize the sign and magnitude inside the
active region.

\subsection{Pauli-blocked activation window and frequency onset}

\begin{figure}
\resizebox{1.0\columnwidth}{!}{
\includegraphics[height=3.6cm]{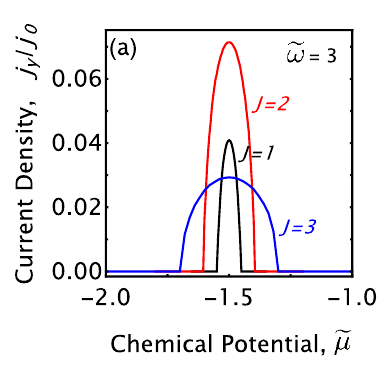}
\includegraphics[height=3.5cm]{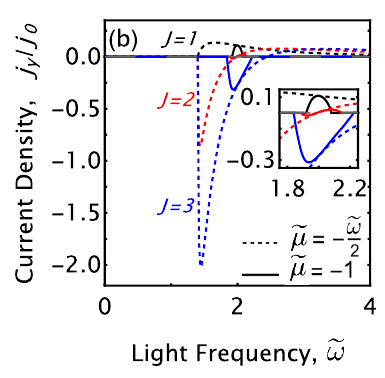}}
\caption{\label{fig:mu-omega}
(a) Dimensionless photon-drag shift current $j_y/j_0$ as a function of the chemical potential $\tilde{\mu}$ for $\tilde{q}=0.1$, $\tilde{\Delta}=0.7$, and $\tilde{\omega}=3$, shown for $J=1,2,3$.
(b) Dimensionless photon-drag shift current $j_y/j_0$ as a function of the light frequency $\tilde{\omega}$ for $\tilde{q}=0.1$ and $\tilde{\Delta}=0.7$. Dashed curves correspond to the representative cut $\tilde{\mu}=-\tilde{\omega}/2$, whereas solid curves correspond to a fixed chemical potential $\tilde{\mu}=-1$. The inset enlarges the response region near $\tilde{\omega}\approx 2$.
Here and in all following figures, $j_0 = e\left(\frac{\pi}{2}\right)\left(\frac{eE}{\hbar\omega}\right)^2 \frac{\varepsilon_0}{\hbar k_0}$.
In the present two-dimensional model, $j_0$ has units of sheet current ($\mathrm{A/m}$). A typical value is $j_0 \approx 3.8\times10^{-13}\,\mathrm{A/m}$ for $E=1\,\mathrm{V/m}$, $\varepsilon_0=\hbar\omega=1\,\mathrm{eV}$, and $k_0=1\,\mathrm{nm}^{-1}$.}
\end{figure}

\begin{figure*}[t]
\resizebox{2.0\columnwidth}{!}{
\includegraphics[height=3.5cm]{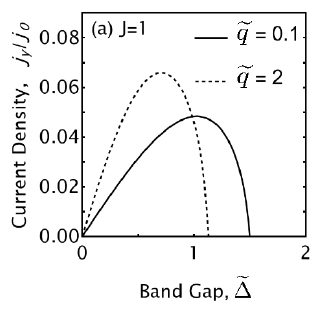}
\includegraphics[height=3.5cm]{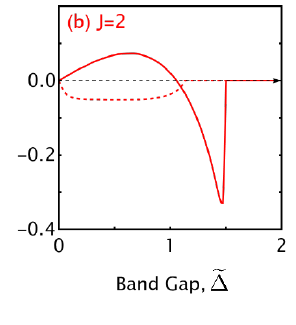}
\includegraphics[height=3.5cm]{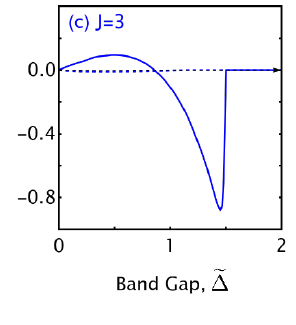}
\includegraphics[height=3.5cm]{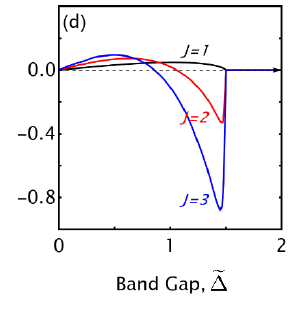}
\includegraphics[height=3.5cm]{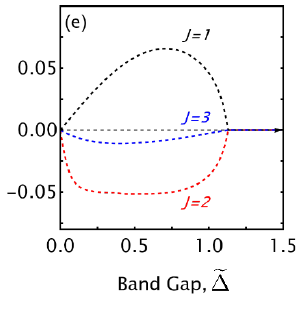}}
\resizebox{2.0\columnwidth}{!}{
\includegraphics[height=3.5cm]{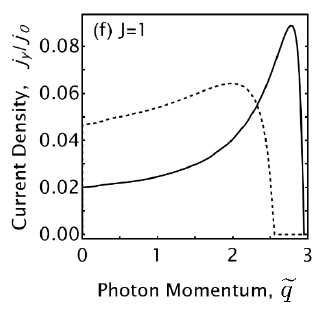}
\includegraphics[height=3.5cm]{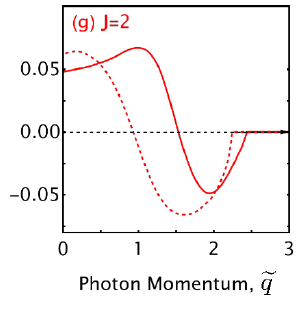}
\includegraphics[height=3.5cm]{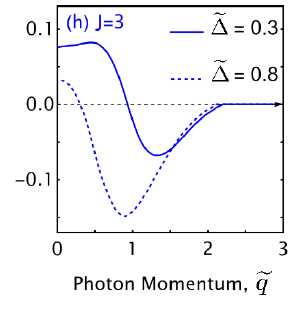}
\includegraphics[height=3.5cm]{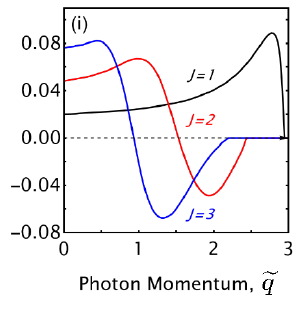}
\includegraphics[height=3.5cm]{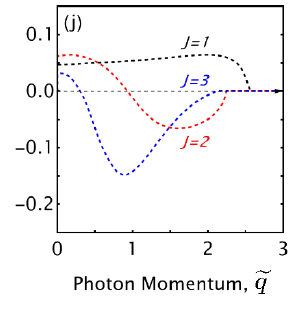}}
\caption{\label{fig:q-delta}
(a--c) Dimensionless photon-drag shift current $j_y/j_0$ as a function of the band gap $\tilde{\Delta}$ for
(a) $J=1$, (b) $J=2$, and (c) $J=3$.
Solid and dashed curves correspond to $\tilde q=0.1$ and $\tilde q=2$, respectively.
(d,e) Comparison of $J=1,2,3$ at fixed photon momentum:
(d) $\tilde q=0.1$ and (e) $\tilde q=2$.
(f--h) Dimensionless photon-drag shift current $j_y/j_0$ as a function of the photon momentum $\tilde q$ for
(f) $J=1$, (g) $J=2$, and (h) $J=3$.
Solid and dashed curves correspond to $\tilde{\Delta}=0.3$ and $\tilde{\Delta}=0.8$, respectively.
(i,j) Comparison of $J=1,2,3$ at fixed band gap:
(i) $\tilde{\Delta}=0.3$ and (j) $\tilde{\Delta}=0.8$.
Unless otherwise stated, the parameters are fixed at $\tilde{\mu}=-1.5$ and $\tilde{\omega}=3$.
}
\end{figure*}

Having established the symmetry and kinematic constraints, we next examine how the finite-$q$ photon-drag shift current is activated as the chemical potential $\mu$ and photon frequency $\omega$ are varied.
The phase-space factor \(\rho(\mathbf{k},\mathbf{q})\) restricts the response to those portions of the energy-conserving contour for which the initial valence state is occupied and the final conduction state is empty.
Consequently, before the chirality-dependent structure of the integrand becomes relevant, the photocurrent is already confined to a restricted window in chemical potential and frequency.

\begin{figure*}[t]
\resizebox{1.7\columnwidth}{!}{
\includegraphics[height=3.5cm]{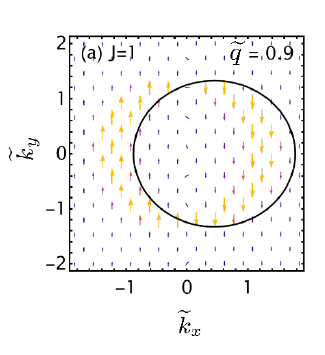}
\includegraphics[height=3.5cm]{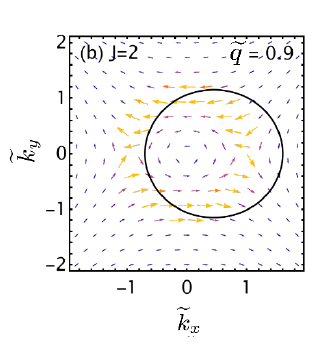}
\includegraphics[height=3.5cm]{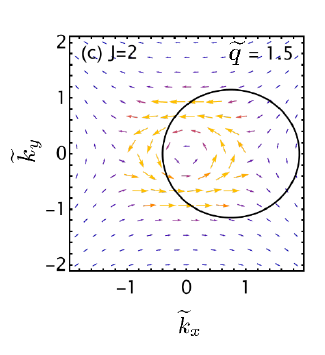}}
\resizebox{1.7\columnwidth}{!}{
\includegraphics[height=3.5cm]{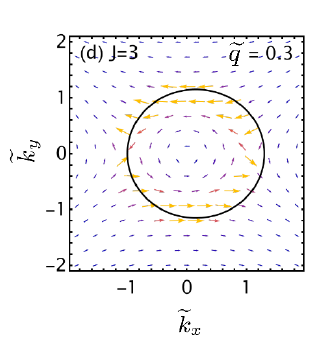}
\includegraphics[height=3.5cm]{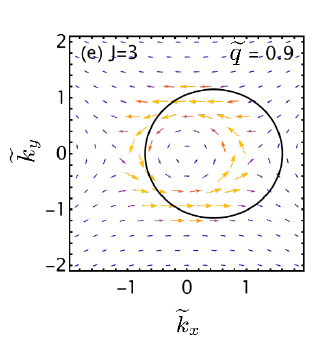}
\includegraphics[height=3.5cm]{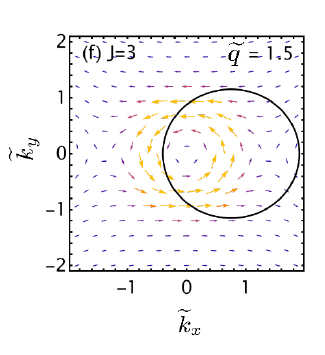}}
\caption{\label{fig:rhoR}
(a--f) Momentum-space vector field of the weighted integrand
$\rho(\mathbf{k},\mathbf{q})\,|\nu(\mathbf{k},\mathbf{q})|^2\,\mathbf r(\mathbf{k},\mathbf{q})$
in the $(\tilde{k}_x,\tilde{k}_y)$ plane, illustrating the microscopic origin of sign reversal.
Panel (a) corresponds to $J=1$ at $\tilde q=0.9$;
panels (b) and (c) correspond to $J=2$ at $\tilde q=0.9$ and $\tilde q=1.5$, respectively;
and panels (d)--(f) correspond to $J=3$ at $\tilde q=0.3$, $\tilde q=0.9$, and $\tilde q=1.5$, respectively.
The black circle marks the chemical-potential boundary whose center is shifted by $\tilde q/2$, and the highlighted yellow/orange segment marks the energy-conserving contour (EC).
Only the EC segment lying in the Pauli-allowed region contributes to the photocurrent.
The contrast between the comparatively simple $J=1$ texture and the more intricate higher-chirality textures shows how varying $\tilde q$ reweights competing sectors of the integrand and can thereby reverse the sign of the net current.
Here $\tilde{\Delta}=0.7$, $\tilde{\mu}=-1.5$, and $\tilde{\omega}=3$.
}
\end{figure*}

This expectation is borne out by the chemical-potential dependence shown in Fig.~\ref{fig:mu-omega}(a).
For all three chirality sectors, the response remains confined to a restricted interval of $\mu$, outside which the photocurrent rapidly collapses.
This localization follows from the occupation factor in \(\rho(\mathbf{k},\mathbf q)\), which selects occupied-to-empty transitions along the resonance contour.
The photocurrent is therefore strongest when this contour overlaps most efficiently with the Pauli-allowed sector.
For the present parameter set, this produces a pronounced maximum near $\tilde \mu=-\tilde\omega/2$, whereas the response is strongly suppressed on either side once the available phase space is depleted.
At the same time, Fig.~\ref{fig:mu-omega}(a) shows that increasing chirality does not lead to a monotonic enhancement of the peak current.
Instead, the main systematic trend is that the interval of chemical potential over which $j_y$ remains nonzero broadens with increasing $J$, while the peak magnitude is reorganized nonmonotonically across the three chirality sectors.

A closely related activation process appears in the frequency dependence shown in Fig.~\ref{fig:mu-omega}(b).
To isolate the role of chemical-potential tuning, we compare a near-optimal Pauli-window cut \(\tilde{\mu}=-\tilde{\omega}/2\) (dashed curves), motivated by Fig.~\ref{fig:mu-omega}(a), with a fixed-chemical-potential cut
\(\tilde{\mu}=-1\) (solid curves).
For sufficiently small \(\tilde{\omega}\), the resonance contour is absent and the photocurrent is negligible.
Once \(\tilde{\omega}\) exceeds the kinematic threshold of Eq.~\eqref{eq:kin}, the resonance contour opens; however, a finite current is generated only from the portion of this contour that also lies in the Pauli-allowed sector.
Along \(\tilde{\mu}=-\tilde{\omega}/2\), the dashed curves follow a near-optimal Pauli window while for the fixed \(\tilde{\mu}=-1\) scan, Pauli blocking restricts the response to a narrower frequency range, with the appreciable signal concentrated near \(\tilde{\omega}\simeq 2\).

Figure~\ref{fig:mu-omega}(b) also shows that chirality affects more than the overall magnitude of the activated response. 
Along the near-optimal cut, the
\(J=1\) photocurrent develops in a comparatively simple manner after the
onset threshold is crossed, whereas the \(J=2\) and \(J=3\) sectors exhibit a
richer frequency evolution, including a pronounced negative dip near onset.
A similar trend is retained for the fixed \(\tilde{\mu}=-1\) scan, although the response is then confined to a narrower frequency window.
The comparison therefore shows that higher chirality already reorganizes the internal sign
structure of the finite-\(q\) photocurrent once the resonance and Pauli constraints activate the response.

\subsection{Chirality-controlled photocurrent sign }
\label{sec:results-C}

The activation window discussed above is common to all chirality sectors, but the sign structure inside that window depends strongly on \(J\).
This distinction is already visible in Fig.~\ref{fig:mu-omega}(b).
Along the representative cut \(\tilde{\mu}=-\tilde{\omega}/2\), the \(J=1\) photocurrent retains the same overall sign across the main resonant interval, whereas the \(J=2\) and \(J=3\) sectors exhibit a pronounced negative lobe near onset followed by a recovery toward smaller positive values at higher frequency.
A similar tendency is retained for the fixed-\(\tilde{\mu}=-1\) scan, although the response is confined to a narrower frequency window.
Thus higher chirality does not merely rescale the response amplitude, but reorganizes the internal sign structure within the kinematically allowed region.

The same conclusion becomes even clearer in the gap dependence shown in Fig.~\ref{fig:q-delta}(a--e).
The current vanishes at $\Delta=0$, where the generalized shift vector $\mathbf{r}(\mathbf{k},\mathbf{q})$ itself disappears, and becomes finite only at nonzero gap.
For $J=1$, the response remains positive once activated and ultimately vanishes only when the energy-conserving contour ceases to exist.
For $J=2$ and $J=3$, however, increasing $\Delta$ drives the photocurrent through zero before the outer cutoff in Eq.~\eqref{eq:kin} is reached.
The zero crossing therefore occurs inside the active region and cannot be identified with the trivial extinction boundary set by the loss of resonance.

\begin{figure*}[t]
\resizebox{1.7\columnwidth}{!}{
\includegraphics[height=3.5cm]{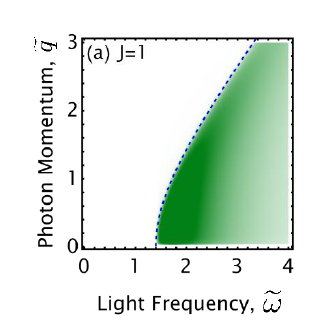}
\includegraphics[height=3.5cm]{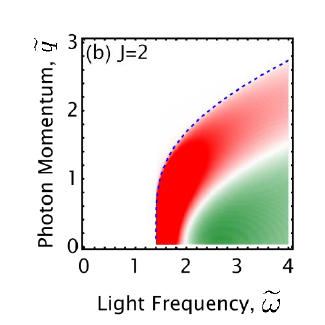}
\includegraphics[height=3.5cm]{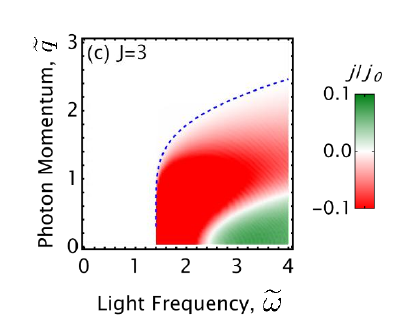}}
\caption{\label{fig:qomega}
(a--c) Phase diagram of the normalized photon-drag shift current $j_y/j_0$ in the $(\tilde{\omega},\tilde q)$ plane for
(a) $J=1$, (b) $J=2$, and (c) $J=3$, at fixed $\tilde{\Delta}=0.7$ and $\tilde{\mu}=-\tilde{\omega}/2$.
Green and red denote positive and negative values of $j_y/j_0$, respectively;
the darkest green corresponds to $j_y/j_0\ge 0.1$, while the darkest red corresponds to $j_y/j_0\le -0.1$.
The dashed blue curve marks the outer kinematic cutoff determined by Eq.~\eqref{eq:kin}.
While the $J=1$ sector remains essentially single-sign throughout the active region, the $J=2$ and $J=3$ sectors develop an internal sign-reversal boundary, shown as a white zero-current contour, across which the photocurrent changes sign while remaining finite on either side within the kinematically allowed finite-$q$ region.
}
\end{figure*}

A closely related pattern appears in the momentum-transfer dependence shown in Fig.~\ref{fig:q-delta}(f--j).
In all cases, the photon-drag shift current vanishes at strictly vertical transition, $q=0$, consistent with the symmetry argument of Sec.~\ref{sec:results-A}.
For finite but arbitrarily small $q$, however, the response approaches the one-sided limit (see Appendix~\ref{app:small_q_jump})
\begin{equation}
\lim_{q\to0^+} j_y(q)
=
\frac{e^2 \varepsilon_0 J}{8\pi \hbar^2 k_0}
\left(\frac{eE}{\hbar\omega}\right)^2
\,
\tilde{\Delta}\,
\frac{
1-J\left(2\tilde{\Delta}/\tilde{\omega}\right)^2
}{
\left[(\tilde{\omega}/2)^2-\tilde{\Delta}^2\right]^{1/2J}
}.
\label{eq:qlimit}
\end{equation}
This singular one-sided limit cannot be captured within the linear-in-$q$ dipole approximation \cite{Likun2021}.
Moreover, Eq.~\eqref{eq:qlimit} already anticipates the higher-chirality sign reversal through the factor
$1-J(2\tilde{\Delta}/\tilde{\omega})^2$.
As $q$ increases further, the $J=1$ response remains sign-stable across the active interval before disappearing at large $q$ owing to the loss of the resonance contour.
By contrast, the $J=2$ and $J=3$ sectors develop internal zero crossings and extended negative-response regions prior to the final extinction.
The sign reversal is therefore not a threshold artifact, but a robust feature of the higher-chirality finite-$q$ response beyond the perturbative small-$q$ regime.

The microscopic origin of this behavior is clarified in Fig.~\ref{fig:rhoR}(a--f), which plots the full momentum-space integrand entering Eq.~\eqref{eq:pdsc},
namely the vector field $\rho(\mathbf{k},\mathbf{q})\,|\nu(\mathbf{k},\mathbf{q})|^2\mathbf{r}(\mathbf{k},\mathbf{q})$,
together with the energy-conserving contour and the chemical-potential boundary.
The net photocurrent is controlled not by the generalized shift vector alone,
but by the weighted integrand \(\rho(\mathbf{k},\mathbf{q})|\nu(\mathbf{k},\mathbf{q})|^2
\mathbf r(\mathbf{k},\mathbf{q})\), evaluated on the resonance contour and
restricted by Pauli blocking.
The black circle marks the chemical-potential boundary, so that only the portion of the contour lying in the Pauli-allowed region contributes to the current.
The sign of the integrated response is therefore controlled by which sectors of the resonant contour survive this selection.

A clear chirality-dependent distinction then emerges.
For $J=1$, the allowed contour samples a comparatively simple and largely co-directional transverse field, so that variations in $\omega$, $\Delta$, or $q$ mainly rescale the response without overturning its sign.
For $J=2$ and $J=3$, by contrast, the weighted integrand develops a much richer texture containing competing positive and negative sectors.
Parameter variation then changes not only the local texture itself, but also how the resonance contour is cut by the chemical-potential boundary, thereby reweighting these competing sectors in a strongly parameter-dependent manner.
The sign reversal for $J\ge2$ is thus produced not by the disappearance of the response, but by a redistribution of weight along a still-existing resonant contour.

This microscopic picture unifies the one-parameter scans in Figs.~\ref{fig:mu-omega} and \ref{fig:q-delta}.
For $J=1$, the allowed contour continues to sample a largely co-directional transverse field as $\omega$, $\Delta$, or $q$ is varied, so the response remains broadly sign-stable.
For $J=2$ and $J=3$, by contrast, the same parameter changes reweight competing positive and negative sectors of the weighted integrand, allowing the net photocurrent to pass through zero and reverse sign within the still-active region.

\subsection{Phase diagram: extinction boundary versus sign-reversal boundary}
\label{sec:results-D}

We now summarize the global sign topology of the finite-$q$ photon-drag shift current in the $(\tilde{\omega},\tilde{q})$ phase diagrams shown in Fig.~\ref{fig:qomega}(a--c), evaluated at fixed $\tilde{\Delta}=0.7$ and along the representative near-optimal cut $\tilde{\mu}=-\tilde{\omega}/2$.
As discussed in Fig.~\ref{fig:mu-omega}(b), this choice provides a convenient benchmark that tracks the activated response most clearly, whereas a fixed-$\tilde{\mu}$ scan retains the same activation physics within a narrower frequency window.
This representation makes it possible to distinguish two physically distinct boundaries:
an outer extinction boundary, beyond which no photocurrent can exist, and an internal sign-reversal boundary, visualized as a zero-current contour, across which the sign of the response changes while the current remains finite on either side.

The outer boundary, shown as the dashed curve in Fig.~\ref{fig:qomega}, is determined by the kinematic cutoff relation in Eq.~\eqref{eq:kin}.
It therefore sets the maximal range of $\omega$ and $q$ over which a resonant non-vertical interband transition can survive at fixed $\Delta$.
This curve should be interpreted as a true extinction boundary:
beyond it, the energy-conserving contour disappears and the photocurrent must vanish identically.
The absence of current in this regime is therefore not caused by cancellation between positive and negative contributions, but simply by the loss of resonant phase space itself.

The internal sign-reversal boundary, shown as the white curve separating positive and negative regions in Fig.~\ref{fig:qomega}, has a different
origin: it is the locus where the integrated current vanishes although the
resonance contour still exists.
Crossing this contour reverses the balance
between positive and negative sectors of the weighted integrand, so the current
changes sign without undergoing kinematic extinction.

To make this distinction more explicit, we denote by $\tilde{\omega}_z$ the dimensionless frequency at which the integrated photocurrent vanishes along the internal boundary.
Although the full zero-current contour is obtained numerically, its small-$q$ anchor is fixed analytically as
\begin{equation}
\left.\tilde{\omega}_z\right|_{q\to0^+}
=
2\sqrt{J}\,\tilde{\Delta},
\label{eq:whitelimit}
\end{equation}
which determines the location of the sign-reversal boundary near the vertical-transition limit.

The comparison among the three chirality sectors makes the physical trend particularly clear.
For $J=1$, the active region remains comparatively simple and is dominated by a single sign over essentially the entire allowed $(\omega,q)$ domain.
For $J=2$ and $J=3$, by contrast, the phase diagram develops a pronounced internal structure, with distinct positive- and negative-response sectors separated by the zero-current contour inside the kinematically allowed region.
Increasing chirality therefore does not merely deform the outer kinematic boundary, but qualitatively reorganizes the interior sign topology of the active region itself before the resonance contour disappears.
Consistent with the one-parameter cuts discussed above, the negative-response sector becomes increasingly prominent as the chirality increases.

The phase diagram therefore provides the most compact synthesis of the full finite-$q$ problem studied here.
Symmetry constrains the photocurrent to be transverse, resonance kinematics determines where a response can exist, and chirality governs how the sign of that response is organized within the allowed region.

\section{Conclusion}

We have investigated the photon-drag shift current in an isotropic single-valley two-dimensional massive chiral Dirac model with chirality index \(J=1,2,3\), by directly evaluating the full finite-\(q\) non-vertical response.
For the geometry considered here, symmetry constrains the photocurrent to be transverse to the photon momentum, and the response vanishes in the strictly vertical-transition limit. 
The main result is that chirality controls not only the magnitude but also the sign structure of the finite-\(q\) response: while the \(J=1\) sector remains broadly sign-stable, the \(J\ge2\) sectors develop internal zero-current
contours and sign reversals within the kinematically allowed domain.

These sign reversals originate from the way in which the resonance contour and the Pauli-allowed sector sample competing positive and negative regions of the finite-\(q\) shift-current integrand. 
They therefore reflect an internal redistribution of geometric weight along a
still-existing resonant contour, rather than the complete extinction of the response, which is instead controlled by the kinematic cutoff at large \(\Delta\) or \(q\). 
The resulting zero-current contour defines a diagnostic boundary between opposite-response sectors and suggests a route to controlling the photocurrent direction by tuning \(\omega\), \(\mu\), \(\Delta\), or \(q\), without changing the driving geometry.

These results establish the isotropic massive chiral Dirac model as a controlled minimal setting for studying chirality-dependent finite-\(q\) shift currents. 
Although the calculation is formulated for a single valley, a time-reversal-related isotropic two-valley completion gives identical
contributions from \(K\) and \(K'\).
Trigonal warping and other anisotropic lattice-scale terms are therefore not required to avoid valley cancellation, but instead generate valley-dependent corrections to the additive isotropic limit. 
Natural extensions include analyzing these anisotropic corrections explicitly, testing the robustness of the sign-reversal structure against finite temperature and disorder broadening, and generalizing the analysis to
three-dimensional chiral systems, including multi-Weyl semimetals.

\begin{acknowledgments}
We thank Dr.~Angga Dito Fauzi, Muhammad Hafidz, and Adiva Nuris Safira for insightful discussions and helpful suggestions.
R.~is supported by a research assistantship from the National Talent Management System at BRIN.
\end{acknowledgments}

\clearpage
\onecolumngrid
\appendix

\section{Derivation of transverse response in Eq. \eqref{eq:symmetryform}}
\label{app:symmetryform}

In this Appendix, we show how Eq. \eqref{eq:symmetryform} follows from the finite-$\mathbf q$ shift-current formula in Eq. \eqref{eq:pdsc} for the isotropic massive chiral Dirac model considered in the main text.
We first write the Hamiltonian in Eq. \eqref{eq:Hk_chiral} in the $d$-vector form
\begin{equation}
H(\mathbf{k})= d_x(\mathbf{k})\sigma_x+d_y(\mathbf{k})\sigma_y+\Delta \sigma_z-\mu\sigma_0,
\end{equation}
with
\begin{equation}
d_x(\mathbf{k})=\varepsilon_0 \left(\frac{|\mathbf{k}|}{k_0}\right)^J \cos(J\phi),\qquad
d_y(\mathbf{k})=\varepsilon_0 \left(\frac{|\mathbf{k}|}{k_0}\right)^J \sin(J\phi),
\end{equation}
where
$\phi=\arg(k_x+i k_y)$.
Under the reflection $\bar{\mathbf{k}}\equiv (k_x,-k_y)$, the coefficients satisfy
\begin{equation}
d_x(\bar{\mathbf{k}})=d_x(\mathbf{k}),\qquad
d_y(\bar{\mathbf{k}})=-d_y(\mathbf{k}),
\end{equation}
so that
\begin{equation}
H(\bar{\mathbf{k}})=H^{*}(\mathbf{k}).
\label{eq:appendix_Hreflection}
\end{equation}

We now choose $\mathbf{q}=q\,\hat{\mathbf{x}}$, as in the main text.
Because the dispersion is isotropic, $\varepsilon_{\pm}(\mathbf{k})$ depends only on
$|\mathbf{k}|$, and therefore
\begin{equation}
\left|\bar{\mathbf{k}}\pm \frac{\mathbf{q}}{2}\right|=
\left|\mathbf{k} \pm \frac{\mathbf{q}}{2}\right|.
\end{equation}
It follows immediately that the phase-space factor in Eq. \eqref{eq:pdsc},
\begin{equation}
\rho(\mathbf{k},\mathbf{q})
=
\left[
f(\varepsilon_{-,\mathbf{k}-\mathbf{q}/2})
-
f(\varepsilon_{+,\mathbf{k}+\mathbf{q}/2})
\right]
\delta\!\left(
\omega_{+,\mathbf{k}+\mathbf{q}/2}
-
\omega_{-,\mathbf{k}-\mathbf{q}/2}
-\omega
\right),
\end{equation}
is even under $k_y\to -k_y$, which implies
\begin{equation}
\rho(\bar{\mathbf{k}},\mathbf{q})=\rho(\mathbf{k},\mathbf{q}).
\label{eq:appendix_rho_even}
\end{equation}

Next, Eq. \eqref{eq:appendix_Hreflection} allows us to choose a gauge such that
\begin{equation}
|u_{c,v}(\bar{\mathbf k})\rangle = |u_{c,v}(\mathbf k)\rangle^{*}.
\label{eq:appendix_gauge}
\end{equation}
From the definition of the Berry connection, $\mathbf{A}_n(\mathbf k)=
i\langle u_n(\mathbf k)|\nabla_{\mathbf k}u_n(\mathbf k)\rangle,$ one then finds the parity relations
\begin{equation}
A_{n,x}(\bar{\mathbf k})=-A_{n,x}(\mathbf k),\qquad
A_{n,y}(\bar{\mathbf k})= A_{n,y}(\mathbf k).
\label{eq:appendix_Berry_parity}
\end{equation}
For light polarized along $\hat{\mathbf x}$, by using Eq. \eqref{eq:appendix_Hreflection}, the interband velocity matrix element in Eq. \eqref{eq:v-matrix} satisfies
\begin{equation}
\nu(\bar{\mathbf k},\mathbf q)=\nu^{*}(\mathbf k,\mathbf q).
\end{equation}
Hence
\begin{equation}
|\nu(\bar{\mathbf k},\mathbf q)|^2=|\nu(\mathbf k,\mathbf q)|^2,
\qquad
\arg \nu(\bar{\mathbf k},\mathbf q)=-\arg \nu(\mathbf k,\mathbf q).
\label{eq:appendix_nu_parity}
\end{equation}
Taking the gradient with respect to the momentum then gives
\begin{equation}
\bigl[\nabla_{\mathbf k}\arg \nu(\bar{\mathbf k},\mathbf q)\bigr]_x
=
-\bigl[\nabla_{\mathbf k}\arg \nu(\mathbf k,\mathbf q)\bigr]_x,
\qquad
\bigl[\nabla_{\mathbf k}\arg \nu(\bar{\mathbf k},\mathbf q)\bigr]_y
=
\phantom{-}\bigl[\nabla_{\mathbf k}\arg \nu(\mathbf k,\mathbf q)\bigr]_y.
\label{eq:appendix_phasegrad_parity}
\end{equation}
Substituting Eqs.~\eqref{eq:appendix_Berry_parity} and
\eqref{eq:appendix_phasegrad_parity} into the generalized shift vector in Eq. \eqref{eq:sv}, we find
\begin{equation}
r_x(\bar{\mathbf k},\mathbf q)=-r_x(\mathbf k,\mathbf q),\qquad
r_y(\bar{\mathbf k},\mathbf q)= r_y(\mathbf k,\mathbf q).
\label{eq:appendix_r_parity}
\end{equation}

We now insert these parity properties into Eq. \eqref{eq:pdsc}. Using Eqs.~\eqref{eq:appendix_rho_even}, \eqref{eq:appendix_nu_parity} and \eqref{eq:appendix_r_parity},
the longitudinal component is odd under $k_y\to -k_y$,
\begin{equation}
\rho(\bar{\mathbf k},\mathbf q)\,|\nu(\bar{\mathbf k},\mathbf q)|^2\,r_x(\bar{\mathbf k},\mathbf q)
=
-\rho(\mathbf k,\mathbf q)\,|\nu(\mathbf k,\mathbf q)|^2\,r_x(\mathbf k,\mathbf q),
\end{equation}
and therefore vanishes after integration over the symmetric momentum plane:
\begin{equation}
j_x(\mathbf q)=0.
\end{equation}
By contrast, the transverse component is even,
\begin{equation}
\rho(\bar{\mathbf k},\mathbf q)\,|\nu(\bar{\mathbf k},\mathbf q)|^2\,r_y(\bar{\mathbf k},\mathbf q)
=
\rho(\mathbf k,\mathbf q)\,|\nu(\mathbf k,\mathbf q)|^2\,r_y(\mathbf k,\mathbf q),
\end{equation}
so that, in general,
\begin{equation}
j_y(\mathbf q)\neq 0.
\end{equation}
Thus, for $\mathbf q=q\,\hat{\mathbf x}$,
\begin{equation}
\mathbf j(\mathbf q)=j_T(q)\,\hat{\mathbf y},
\label{eq:appendix_qx}
\end{equation}
with $j_T(q)$ a scalar function.

Finally, because the continuum model is rotationally symmetric, rotating both $\mathbf k$ and $\mathbf q$ by an in-plane angle $\theta$
must rotate the current by the same angle.
Therefore, Eq.~\eqref{eq:appendix_qx} generalizes to an arbitrary in-plane photon momentum as
\begin{equation}
\mathbf j(\mathbf q)
=
\alpha(\omega,\Delta,\mu,|\mathbf q|)\,
\hat{\mathbf z}\times \mathbf q,
\end{equation}
which is Eq. \eqref{eq:symmetryform} of the main text.
Here
$\alpha(\omega,\Delta,\mu,|\mathbf q|)$ is a scalar response function that depends only on rotationally invariant quantities.
As an immediate consequence,
\begin{equation}
\mathbf j(-\mathbf q)=-\mathbf j(\mathbf q),
\end{equation}
and hence $\mathbf j(\mathbf q=0)=0$.

\section{Derivation of the kinematic cutoff in Eq.~\eqref{eq:kin}}
\label{app:kinematic_cutoff}

In this Appendix, we derive the kinematic condition that determines the existence of non-vertical resonant interband transitions, which gives the dashed boundary in Fig.~\ref{fig:qomega}.
The starting point is the phase-space factor in Eq.~\eqref{eq:pdsc},
\begin{equation}
\rho(\mathbf{k},\mathbf{q})
=
\Bigl[
f\!\left(\varepsilon_{-,\mathbf{k}-\mathbf{q}/2}\right)
-
f\!\left(\varepsilon_{+,\mathbf{k}+\mathbf{q}/2}\right)
\Bigr]
\delta\!\left(
\omega_{+,\mathbf{k}+\mathbf{q}/2}
-
\omega_{-,\mathbf{k}-\mathbf{q}/2}
-
\omega
\right),
\label{eq:B1}
\end{equation}
which shows that the photocurrent can be nonzero only when the energy-conserving condition
\begin{equation}
\hbar\omega
=
\varepsilon_{+}\!\left(\mathbf{k}+\frac{\mathbf{q}}{2}\right)
-
\varepsilon_{-}\!\left(\mathbf{k}-\frac{\mathbf{q}}{2}\right)
\label{eq:B2}
\end{equation}
has at least one solution for some momentum \(\mathbf{k}\).

Using the dispersion in Eq.~\eqref{eq:energy}, and introducing the dimensionless variables
\(\tilde{\mathbf{k}}=\mathbf{k}/k_0\), \(\tilde q=q/k_0\), and \(\tilde\omega=\hbar\omega/\varepsilon_0\),
with \(q=|\mathbf q|\), Eq.~\eqref{eq:B2} becomes
\begin{equation}
\tilde{\omega}
=
\sqrt{\left|\tilde{\mathbf{k}}+\frac{\tilde q}{2}\hat{\mathbf{x}}\right|^{2J}+\tilde{\Delta}^{2}}
+
\sqrt{\left|\tilde{\mathbf{k}}-\frac{\tilde q}{2}\hat{\mathbf{x}}\right|^{2J}+\tilde{\Delta}^{2}}.
\label{eq:B3}
\end{equation}
As expected, the chemical potential drops out of the resonance condition itself, so the kinematic cutoff is determined solely by \(\tilde\omega\), \(\tilde\Delta\), and \(\tilde q\).
The role of \(\tilde\mu\) is only to further restrict the allowed transitions through Pauli blocking.

To determine when Eq.~\eqref{eq:B3} admits a solution, we seek the minimum possible transition energy at fixed \(\tilde q\).
Define
\begin{equation}
a=
\left|\tilde{\mathbf{k}}+\frac{\tilde q}{2}\hat{\mathbf{x}}\right|,
\qquad
b=
\left|\tilde{\mathbf{k}}-\frac{\tilde q}{2}\hat{\mathbf{x}}\right|,
\label{eq:B4}
\end{equation}
so that
\begin{equation}
\tilde{\omega}
=
f(a)+f(b),
\qquad
f(x)=\sqrt{x^{2J}+\tilde{\Delta}^{2}}.
\label{eq:B5}
\end{equation}

For \(x\ge0\), the function \(f(x)\) is monotonically increasing,
\begin{equation}
f'(x)
=
\frac{Jx^{2J-1}}{\sqrt{x^{2J}+\tilde{\Delta}^{2}}}
>0,
\label{eq:B6}
\end{equation}
and also convex,
\begin{equation}
f''(x)
=
\frac{
Jx^{2J-2}\left[(2J-1)\tilde{\Delta}^{2}+(J-1)x^{2J}\right]
}{
\left(x^{2J}+\tilde{\Delta}^{2}\right)^{3/2}
}
>0.
\label{eq:B7}
\end{equation}

We now use two simple inequalities.
First, by the triangle inequality,
\begin{equation}
a+b
=
\left|\tilde{\mathbf{k}}+\frac{\tilde q}{2}\hat{\mathbf{x}}\right|
+
\left|\tilde{\mathbf{k}}-\frac{\tilde q}{2}\hat{\mathbf{x}}\right|
\ge
\tilde q.
\label{eq:B8}
\end{equation}
Second, because \(f(x)\) is convex, Jensen's inequality gives
\begin{equation}
f(a)+f(b)
\ge
2f\!\left(\frac{a+b}{2}\right).
\label{eq:B9}
\end{equation}
Since \(f(x)\) is increasing and \((a+b)/2\ge \tilde q/2\), we obtain
\begin{equation}
f(a)+f(b)
\ge
2f\!\left(\frac{a+b}{2}\right)
\ge
2f\!\left(\frac{\tilde q}{2}\right).
\label{eq:B10}
\end{equation}

Therefore, the minimum transition energy is
\begin{equation}
\tilde{\omega}_{\min}
=
2\sqrt{\left(\frac{\tilde q}{2}\right)^{2J}+\tilde{\Delta}^{2}}.
\label{eq:B11}
\end{equation}
The lower bound is saturated when \(a=b=\tilde q/2\), which is realized by the symmetric configuration \(\tilde{\mathbf{k}}=0\).
Physically, this means that the lowest-energy non-vertical transition occurs when the initial and final states are located symmetrically at \(\pm \tilde q/2\), i.e., as close as possible to the band edge while maintaining the required momentum separation.

Hence, a resonant interband transition exists only when
\begin{equation}
\tilde{\omega}
\ge
2\sqrt{\tilde{\Delta}^{2}+\left(\frac{\tilde q}{2}\right)^{2J}}.
\label{eq:B12}
\end{equation}
Squaring both sides gives
\begin{equation}
\tilde{\omega}^{2}
\ge
(2\tilde{\Delta})^{2}
+
\frac{\tilde q^{\,2J}}{2^{\,2J-2}}
=
(2\tilde{\Delta})^{2}
+
\left(\frac{\tilde q^{\,J}}{2^{J-1}}\right)^{2},
\label{eq:B13}
\end{equation}
which is Eq.~\eqref{eq:kin} of the main text.

Equation~\eqref{eq:B13} therefore defines the kinematic cutoff for the non-vertical interband process.
Below this boundary, the delta function in Eq.~\eqref{eq:B1} has no support, so the energy-conserving contour is absent and the photocurrent vanishes.
Above it, resonant transitions become allowed and the photon-drag shift current can be activated, subject to the additional Pauli-blocking constraint set by the chemical potential.

\section{Small-\(q\) analysis of the sign-reversal boundary}
\label{app:smallq_sign}

In Fig.~\ref{fig:qomega}(b,c), the white curve separating the positive and negative regions does not represent the kinematic cutoff of Eq.~\eqref{eq:kin}.
Instead, it is the locus at which the transverse photon-drag shift current changes sign.
At finite \(q\), this sign-reversal boundary is defined implicitly by Eq.~\eqref{eq:pdsc} and, in general, does not reduce to a simple closed-form expression analogous to Eq.~\eqref{eq:kin}.
Nevertheless, its origin can be understood analytically in the small-\(q\) limit, where one can derive the onset of the white curve and explain why it appears only for \(J\ge2\).

For \(x\)-polarized light, the \(q\to0\) limit of Eq.~\eqref{eq:v-matrix} reduces to the vertical-transition interband velocity matrix element \(\nu_x(\mathbf{k})\).
To evaluate it, we write \(\mathbf{k}=k(\cos\phi,\sin\phi)\) and introduce
\begin{equation}
\tilde E_k=\sqrt{\tilde k^{\,2J}+\tilde\Delta^2},
\qquad
\alpha=\frac{\tilde\Delta}{\tilde E_k},
\label{eq:C1def}
\end{equation}
so that the conduction- and valence-band spinors of the massive chiral Dirac Hamiltonian in Eq.~\eqref{eq:Hk_chiral} may be written as
\begin{equation}
|u_c\rangle=
\begin{pmatrix}
\cos(\theta/2)\\
e^{iJ\phi}\sin(\theta/2)
\end{pmatrix},
\qquad
|u_v\rangle=
\begin{pmatrix}
-\,e^{-iJ\phi}\sin(\theta/2)\\
\cos(\theta/2)
\end{pmatrix},
\label{eq:C1}
\end{equation}
with
\begin{equation}
\cos\theta=\alpha,
\qquad
\sin\theta=\sqrt{1-\alpha^2}.
\label{eq:C1b}
\end{equation}

The vertical-transition interband velocity matrix element is then
\begin{equation}
\nu_x(\mathbf{k})
=
\frac{\varepsilon_0 J}{\hbar k_0}\,
\tilde k^{\,J-1}
e^{-iJ\phi}
\left(\alpha\cos\phi+i\sin\phi\right),
\label{eq:C2}
\end{equation}
so that
\begin{equation}
|\nu_x(\mathbf{k})|^2
=
\left(\frac{\varepsilon_0 J}{\hbar k_0}\right)^2
\tilde k^{\,2J-2}
\left(\alpha^2\cos^2\phi+\sin^2\phi\right).
\label{eq:C3}
\end{equation}

Next, we evaluate the \(y\)-component of the shift vector in the \(q\to0\) limit.
Using Eq.~\eqref{eq:sv}, one finds
\begin{equation}
r_y^{(0)}(\mathbf{k})
=
\frac{\alpha\cos\phi}{\tilde k}\,
\frac{J\alpha^2-1}{\alpha^2\cos^2\phi+\sin^2\phi}.
\label{eq:C4}
\end{equation}

Multiplying Eqs.~\eqref{eq:C3} and \eqref{eq:C4} gives the gauge-invariant integrand factor
\begin{equation}
|\nu_x(\mathbf{k})|^2\,r_y^{(0)}(\mathbf{k})
=
\left(\frac{\varepsilon_0 J}{\hbar k_0}\right)^2
\tilde k^{\,2J-3}\,
\alpha\left(J\alpha^2-1\right)\cos\phi.
\label{eq:C5}
\end{equation}
Thus, apart from an overall positive prefactor, the sign structure is controlled by the factor \(J\alpha^2-1\).

We now combine this result with the Pauli-blocking factor in Eq.~\eqref{eq:pdsc}.
In Fig.~\ref{fig:qomega}, we set \(\tilde\mu=-\tilde\omega/2\).
For small but finite \(q\parallel\hat{\mathbf x}\), the resonance condition approaches the vertical-transition limit
\begin{equation}
2\tilde E_k=\tilde\omega,
\label{eq:C6}
\end{equation}
so that the resonant contour is approximately a circle.

Expanding the initial valence-band energy for the non-vertical process gives
\begin{equation}
\varepsilon_{-}\!\left(\mathbf{k}-\frac{\mathbf q}{2}\right)
\simeq
\varepsilon_0
\left(
-\tilde E_k+\frac{\tilde q}{2}\frac{\partial \tilde E_k}{\partial \tilde k_x}
\right)
-\mu.
\label{eq:C7}
\end{equation}
Substituting \(\tilde\mu=-\tilde\omega/2\) together with Eq.~\eqref{eq:C6}, we obtain
\begin{equation}
\varepsilon_{-}\!\left(\mathbf{k}-\frac{\mathbf q}{2}\right)
\simeq
\varepsilon_0\,
\frac{\tilde q}{2}
\frac{\partial \tilde E_k}{\partial \tilde k_x}.
\label{eq:C8}
\end{equation}
Since
\begin{equation}
\frac{\partial \tilde E_k}{\partial \tilde k_x}
=
\frac{J\tilde k^{\,2J-1}}{\tilde E_k}\cos\phi,
\label{eq:C9}
\end{equation}
the occupied part of the resonant contour at \(T=0\) is selected by
\begin{equation}
\varepsilon_{-}\!\left(\mathbf{k}-\frac{\mathbf q}{2}\right)<0
\quad\Longrightarrow\quad
\cos\phi<0.
\label{eq:C10}
\end{equation}
Therefore, in the small-\(q\) limit the integral receives its dominant contribution only from \(\phi\in[\pi/2,3\pi/2]\).

Using Eq.~\eqref{eq:C5}, the current near \(q\to0^+\) behaves as
\begin{equation}
j_y(q\to0^+)
\propto
\int_{\pi/2}^{3\pi/2}
d\phi\,
|\nu_x(\mathbf{k})|^2\,r_y^{(0)}(\mathbf{k})
\propto
1-J\alpha^2.
\label{eq:C11}
\end{equation}
On the resonant contour, Eq.~\eqref{eq:C6} gives
\begin{equation}
\alpha=\frac{\tilde\Delta}{\tilde E_k}=\frac{2\tilde\Delta}{\tilde\omega},
\label{eq:C12}
\end{equation}
so that
\begin{equation}
j_y(q\to0^+)
\propto
1-J\left(\frac{2\tilde\Delta}{\tilde\omega}\right)^2.
\label{eq:C13}
\end{equation}

We now define \(\tilde\omega_z\) as the dimensionless zero-current frequency, i.e., the frequency at which the integrated photocurrent vanishes along the internal sign-reversal boundary.
The sign-reversal condition is therefore
\begin{equation}
1-J\left(\frac{2\tilde\Delta}{\tilde\omega_z}\right)^2=0,
\label{eq:C14}
\end{equation}
which yields
\begin{equation}
\left.\tilde\omega_z\right|_{q\to0^+}
=
2\sqrt{J}\,\tilde\Delta.
\label{eq:C15}
\end{equation}
This is Eq.~\eqref{eq:whitelimit} of the main text.

Equation~\eqref{eq:C15} gives the analytic starting point of the white curve in Fig.~\ref{fig:qomega}.
It also explains why the sign-reversal boundary is absent for \(J=1\) but appears for \(J=2,3\).
Indeed, the interband onset at \(q\to0\) is \(\tilde\omega=2\tilde\Delta\), while the zero-current line starts at \(\tilde\omega_z=2\sqrt{J}\,\tilde\Delta\).
For \(J=1\), these two values coincide, so the current does not pass through a finite-frequency sign-reversal region after the onset.
By contrast, for \(J\ge2\) one has
\begin{equation}
2\tilde\Delta < 2\sqrt{J}\,\tilde\Delta,
\label{eq:C16}
\end{equation}
which opens a finite frequency window in which the current is allowed but has the opposite sign before crossing the white boundary.

For finite \(\tilde q\), the full sign-reversal boundary is determined implicitly by the zero-current condition
\begin{equation}
j_y(\tilde\omega,\tilde q;\tilde\Delta,\tilde\mu)=0,
\label{eq:C17}
\end{equation}
and therefore does not, in general, admit a simple closed form.
However, Eq.~\eqref{eq:C15} provides its exact small-\(q\) anchor and captures the analytic origin of the white curve seen in Fig.~\ref{fig:qomega}(b,c).

\section{One-sided small-\(q\) limit at \(T=0\) and \(\tilde{\mu}=-\tilde{\omega}/2\)}
\label{app:small_q_jump}

In this Appendix, we derive the finite one-sided limit of the photon-drag shift current in the singular case \(T=0\) and \(\tilde{\mu}=-\tilde{\omega}/2\), for which the current exhibits a discontinuous jump at \(q=0\).

We start from Eq.~\eqref{eq:pdsc} of the main text at \(q=0\), where the transition is vertical and the resonance condition reduces to Eq.~\eqref{eq:C6}, so that the resonant contour is a full circle.
In this limit, the angular dependence of the gauge-invariant integrand is proportional to \(\cos\phi\), and therefore the full angular integral vanishes:
\begin{equation}
j_y(q=0)\propto \int_0^{2\pi}\cos\phi\,d\phi = 0.
\label{eq:D1}
\end{equation}
Hence, the current at exactly \(q=0\) is zero.

The singular behavior arises from the Pauli-blocking factor at \(T=0\).
Since \(f(\varepsilon)=\Theta(-\varepsilon)\), we expand the valence-band energy for small \(q\) on the resonant contour, where
\begin{equation}
2\tilde E_k=\tilde\omega,
\label{eq:D2}
\end{equation}
with \(\tilde E_k=\sqrt{\tilde k^{\,2J}+\tilde\Delta^2}\).
From Eq.~\eqref{eq:C8}, we have
\begin{equation}
\varepsilon_{-}\!\left(\mathbf{k}-\frac{\mathbf q}{2}\right)
\simeq
\varepsilon_0\,
\frac{\tilde q}{2}
\frac{\partial \tilde E_k}{\partial \tilde k_x}.
\label{eq:D3}
\end{equation}
Because the dispersion is isotropic,
\begin{equation}
\frac{\partial \tilde E_k}{\partial \tilde k_x}
=
\frac{d\tilde E_k}{d\tilde k}\cos\phi,
\qquad
\frac{d\tilde E_k}{d\tilde k}>0,
\label{eq:D4}
\end{equation}
the sign of \(\varepsilon_{-,\mathbf{k}-\mathbf{q}/2}\) is determined by \(q\cos\phi\).
Therefore, for \(q\to0^+\),
\begin{equation}
f\!\left(\varepsilon_{-,\mathbf{k}-\mathbf{q}/2}\right)
-
f\!\left(\varepsilon_{+,\mathbf{k}+\mathbf{q}/2}\right)
\longrightarrow
\Theta(-\cos\phi),
\label{eq:D5}
\end{equation}
which means that only the left semicircle contributes.
This is the origin of the jump discontinuity.

In the small-\(q\) limit, all smooth factors may be evaluated at \(q=0\), so Eq.~\eqref{eq:pdsc} becomes
\begin{equation}
\lim_{q\to0^+}j_y(q)
=
C
\int \frac{d^2k}{(2\pi)^2}\,
\Theta(-\cos\phi)\,
\delta(2E_k-\hbar\omega)\,
|\nu_x(\mathbf{k})|^2\,
r_y^{(0)}(\mathbf{k}),
\label{eq:D6}
\end{equation}
where \(r_y^{(0)}(\mathbf{k})\) denotes the \(q=0\) shift vector, and \(\nu_x(\mathbf{k})\) is the vertical-transition interband velocity matrix element defined in Appendix~\ref{app:smallq_sign}.

Using Eq.~\eqref{eq:C5}, we write
\begin{equation}
d^2k = k\,dk\,d\phi = k_0^2\,\tilde k\,d\tilde k\,d\phi,
\label{eq:D7}
\end{equation}
and
\begin{equation}
\delta(2E_k-\hbar\omega)
=
\frac{1}{\varepsilon_0}\,
\delta(2\tilde E_k-\tilde\omega).
\label{eq:D8}
\end{equation}
Then Eq.~\eqref{eq:D6} becomes
\begin{equation}
\lim_{q\to0^+}j_y(q)
=
C
\frac{\varepsilon_0 J^2}{\hbar^2 k_0}
\int \frac{d\phi}{(2\pi)^2}\,
\Theta(-\cos\phi)\cos\phi
\int d\tilde k\,
\tilde k^{\,2J-2}\,
\alpha(J\alpha^2-1)\,
\delta(2\tilde E_k-\tilde\omega).
\label{eq:D9}
\end{equation}

The angular integral is elementary:
\begin{equation}
\int d\phi\,\Theta(-\cos\phi)\cos\phi
=
\int_{\pi/2}^{3\pi/2}\cos\phi\,d\phi
=
-2.
\label{eq:D10}
\end{equation}

Next, define the resonant momentum \(\tilde k_\omega\) by
\begin{equation}
2\tilde E_k=\tilde\omega
\quad\Longrightarrow\quad
\tilde k_\omega
=
\left[
\left(\frac{\tilde\omega}{2}\right)^2-\tilde\Delta^2
\right]^{1/2J},
\label{eq:D11}
\end{equation}
which requires \(\tilde\omega>2\tilde\Delta\).
Since
\begin{equation}
\frac{d\tilde E_k}{d\tilde k}
=
\frac{J\tilde k^{\,2J-1}}{\tilde E_k},
\label{eq:D12}
\end{equation}
the radial delta-function integral gives
\begin{equation}
\int d\tilde k\,
\tilde k^{\,2J-2}\,
\delta(2\tilde E_k-\tilde\omega)
=
\frac{\tilde\omega}{4J\,\tilde k_\omega}.
\label{eq:D13}
\end{equation}

On the resonant contour, we also have \(\alpha=2\tilde\Delta/\tilde\omega\) from Eq.~\eqref{eq:C12}.
Combining Eqs.~\eqref{eq:D9}--\eqref{eq:D13}, we obtain
\begin{equation}
\lim_{q\to0^+}j_y(q)
=
\frac{e^2\varepsilon_0 J}{8\pi\hbar^2 k_0}
\left(\frac{eE}{\hbar\omega}\right)^2
\frac{
\tilde\Delta
\left[
1-J\left(\frac{2\tilde\Delta}{\tilde\omega}\right)^2
\right]
}{
\left[
\left(\frac{\tilde\omega}{2}\right)^2-\tilde\Delta^2
\right]^{1/2J}
},
\qquad
(\tilde\omega>2\tilde\Delta),
\label{eq:D14}
\end{equation}
which is Eq.~\eqref{eq:qlimit} of the main text.

The one-sided limit for \(q\to0^-\) is obtained in the same way, except that the Pauli factor selects the right semicircle,
\(\Theta(\cos\phi)\), which flips the sign of the angular integral.
Therefore,
\begin{equation}
\lim_{q\to0^-}j_y(q)
=
-\lim_{q\to0^+}j_y(q).
\label{eq:D15}
\end{equation}

Thus, the current satisfies
\begin{equation}
j_y(0)=0,
\qquad
\lim_{q\to0^\pm}j_y(q)=\pm j^\ast,
\label{eq:D16}
\end{equation}
with the finite amplitude \(j^\ast\) given by Eq.~\eqref{eq:D14}.

\section{Relation between the two time-reversal valleys}
\label{app:valley-relation}

In the main text we evaluate the finite-\(q\) photon-drag shift current for a single isotropic chiral Dirac valley. 
For completeness, we now clarify how this response is related to that of the time-reversal partner valley. 
This point is important because the oddness of the single-valley response under \(\mathbf q\to-\mathbf q\) should not be confused with a cancellation between the \(K\) and \(K'\) valleys.

We take the valley \(K\) to be described by Eq.~\eqref{eq:Hk_chiral}, namely
\begin{equation}
H_K(\mathbf k)
=
\varepsilon_0
\left[
\left(\frac{k_-}{k_0}\right)^J\sigma_+
+
\left(\frac{k_+}{k_0}\right)^J\sigma_-
\right]
+\Delta\sigma_z-\mu\sigma_0 ,
\label{eq:HK_app}
\end{equation}
where \(k_\pm=k_x\pm i k_y\). 
The corresponding time-reversal partner valley is defined by
\begin{equation}
H_{K'}(\mathbf k)=H_K^*(-\mathbf k).
\label{eq:HKp_TR}
\end{equation}
Equivalently,
\begin{equation}
H_{K'}(\mathbf k)
=
\varepsilon_0
\left[
\left(\frac{-k_x-i k_y}{k_0}\right)^J\sigma_+
+
\left(\frac{-k_x+i k_y}{k_0}\right)^J\sigma_-
\right]
+\Delta\sigma_z-\mu\sigma_0 .
\label{eq:HKp_app}
\end{equation}
Let \(\Theta\) denote the spinless time-reversal operation, which acts as complex conjugation in the present pseudospin basis. 
From Eq.~\eqref{eq:HKp_TR}, the Bloch eigenstates in the two valleys can be chosen such that
\begin{equation}
|u_{n,K'}(\mathbf k)\rangle
=
\Theta |u_{n,K}(-\mathbf k)\rangle
=
|u_{n,K}(-\mathbf k)\rangle^* ,
\label{eq:state_TR}
\end{equation}
up to a smooth gauge phase. 
The corresponding eigenenergies satisfy
\begin{equation}
\varepsilon_{n,K'}(\mathbf k)
=
\varepsilon_{n,K}(-\mathbf k)
=
\varepsilon_n(|\mathbf k|).
\label{eq:energy_TR}
\end{equation}
Thus the two valleys have identical isotropic dispersions, but opposite pseudospin winding.

For a photon carrying in-plane momentum \(\mathbf q\), the finite-\(q\) interband transition in the \(K\) valley is
\begin{equation}
|u_{v,K}(\mathbf k-\mathbf q/2)\rangle
\rightarrow
|u_{c,K}(\mathbf k+\mathbf q/2)\rangle .
\label{eq:transition_K}
\end{equation}
In global momentum coordinates, the corresponding transition in the \(K'\) valley has the same physical momentum transfer,
\begin{equation}
|u_{v,K'}(\mathbf k-\mathbf q/2)\rangle
\rightarrow
|u_{c,K'}(\mathbf k+\mathbf q/2)\rangle .
\label{eq:transition_Kp_global}
\end{equation}
Under time reversal, however, Eq.~\eqref{eq:transition_Kp_global} is mapped to a \(K\)-valley transition with momentum transfer \(-\mathbf q\):
\begin{equation}
|u_{v,K}(-\mathbf k+\mathbf q/2)\rangle
\rightarrow
|u_{c,K}(-\mathbf k-\mathbf q/2)\rangle .
\label{eq:transition_TR}
\end{equation}
Defining \(\boldsymbol{\kappa}=-\mathbf k\), this becomes
\begin{equation}
|u_{v,K}(\boldsymbol{\kappa}-(-\mathbf q)/2)\rangle
\rightarrow
|u_{c,K}(\boldsymbol{\kappa}+(-\mathbf q)/2)\rangle .
\label{eq:transition_K_minus_q}
\end{equation}
Therefore, the \(K'\)-valley process driven by \(\mathbf q\) is the time-reversed counterpart of the \(K\)-valley process driven by \(-\mathbf q\).

The current operator is odd under time reversal,
\begin{equation}
\Theta \hat{\mathbf j}\Theta^{-1}
=
-\hat{\mathbf j}.
\label{eq:current_TR_odd}
\end{equation}
Consequently, the valley-resolved finite-\(q\) currents obey
\begin{equation}
\mathbf j_{K'}(\mathbf q)
=
-\mathbf j_K(-\mathbf q),
\label{eq:valley_current_relation}
\end{equation}
which is Eq. \eqref{eq:valeysrelation} of the main text.
For the isotropic single-valley model considered in the main text, Appendix~\ref{app:symmetryform} shows that the finite-\(q\) current has the transverse form
\begin{equation}
\mathbf j_K(\mathbf q)
=
\alpha(\omega,\Delta,\mu,|\mathbf q|)
\,\hat{\mathbf z}\times \mathbf q ,
\label{eq:jK_transverse_app}
\end{equation}
where \(\alpha\) is a scalar function of rotationally invariant parameters. 
Hence
\begin{equation}
\mathbf j_K(-\mathbf q)
=
-\mathbf j_K(\mathbf q).
\label{eq:jK_odd_app}
\end{equation}
Combining Eqs.~\eqref{eq:valley_current_relation} and \eqref{eq:jK_odd_app}, one obtains
\begin{equation}
\mathbf j_{K'}(\mathbf q)
=
\mathbf j_K(\mathbf q).
\label{eq:valley_additive}
\end{equation}
Thus, in a time-reversal-related isotropic two-valley completion of the present model, the two valley contributions are identical rather than cancelling. 
The total charge current is therefore
\begin{equation}
\mathbf j_{\rm tot}(\mathbf q)
=
\mathbf j_K(\mathbf q)+\mathbf j_{K'}(\mathbf q)
=
2\mathbf j_K(\mathbf q).
\label{eq:total_two_valley}
\end{equation}

\twocolumngrid
\bibliographystyle{apsrev4-2}
\bibliography{ref}

\end{document}